\documentstyle[11pt,citesort]{article}
\input epsf
\def \a {\alpha}
\def \b {\beta}

\def \is {\! & \! = \! & \! }
\newcommand{\newsubsection}[1]{
\vspace{1cm}
\pagebreak[3]
\addtocounter{subsection}{1}
\addcontentsline{toc}{subsection}{\protect
\numberline{%\arabic{section}.
\arabic{subsection}}{#1}}
\noindent{\large\bf %\thesection.
#1}
\nopagebreak
\vspace{2mm}
\nopagebreak}

\newcommand{\figuur}[3]{
\begin{figure}[t]\begin{center}
\leavevmode\hbox{\epsfxsize=#2 \epsffile{#1.eps}}\\[3mm]
\parbox{15.5cm}{\small
\it #3}
\end{center}
\end{figure}}
\newlength{\extraspace}
\newlength{\extraspaces}
\newcommand{\ba}{\begin{eqnarray}
\addtolength{\abovedisplayskip}{\extraspaces}
\addtolength{\belowdisplayskip}{\extraspaces}
\addtolength{\abovedisplayshortskip}{\extraspaces}
\addtolength{\belowdisplayshortskip}{\extraspaces}}
\newcommand{\ea}{\end{eqnarray}}
\newcommand{\be}{\begin{equation}
\addtolength{\abovedisplayskip}{\extraspaces}
\addtolength{\belowdisplayskip}{\extraspaces}
\addtolength{\abovedisplayshortskip}{\extraspace}
\addtolength{\belowdisplayshortskip}{\extraspace}}
\newcommand{\ee}{\end{equation}}
\newcommand{\wilde}{\tilde}
\begin{document}
\addtolength{\baselineskip}{.8mm}
\def\calN{{\cal N}}
\def\Box{\square}
\def \Tr{\mbox{Tr\,}}
\def \tr{\mbox{Tr\,}}
\def \a{\alpha}
\def \da{{\dot a}}
\def \db{{\dot b}}
\def \dc{{\dot c}}
\def \dd{{\dot d}}
\def \CH{\mathfrak {CH}}
\def \b{\beta}
\def \vt{\vartheta}
\def \ttheta{{\tilde \theta}}
\def \tSigma{{\tilde \Sigma}}
\def\ignorethis#1{}
\def\ar#1#2{\begin{array}{#1}#2\end{array}}
\def\bear{\begin{eqnarray}}
\def\eear{\end{eqnarray}}
\def\p{\partial }
\def\IR{{\bf R}}
\def\nul{{\mbox{\tiny $0$}}}
\def \two {{{}_{2}}}

\newcommand{\ggt}{\!>\!\!>\!}
\newcommand{\llt}{\!<\!\!<\!}
%%%%%%%%%%%%%%%%%%%%%%%%
\def\dag{\dagger}
\def\bea{\begin{eqnarray}}
\def\eea{\end{eqnarray}}
\def\appendix#1{
  \addtocounter{section}{1}
  \setcounter{equation}{0}
  \renewcommand{\thesection}{\Alph{section}}
  \section*{Appendix \thesection\protect\indent \parbox[t]{11.15cm}
  {#1} }
  \addcontentsline{toc}{section}{Appendix \thesection\ \ \ #1}
  }

\begin{titlepage}
\begin{center}

{\hbox to\hsize{ \hfill PUPT-2052}}
{\hbox to\hsize{ \hfill
NSF-KITP-02-153}} {\hbox to\hsize{ \hfill hep-th/0210102}}

\bigskip

\vspace{3\baselineskip}

{\LARGE \bf Tracing the String\,:}\\[6mm]
{\Large \bf BMN Correspondence at Finite $J^2/N$}

\bigskip
\bigskip
\bigskip

{\large John Pearson,${}^1$ Marcus Spradlin,${}^1$ Diana Vaman,${}^1$}\\[5mm]

{\large Herman Verlinde ${}^1$ and Anastasia Volovich ${}^2$}\\[1cm]

{${}^1$ \it Department of Physics, Princeton University,
Princeton,
NJ 08544}\\[6mm]

{${}^2$ \it Kavli Institute for Theoretical Physics, Santa
Barbara, CA 93106}

\vspace*{1.5cm}

{\bf Abstract}\\
\end{center}
Employing the string bit formalism of hep-th/0209215, we identify
the basis transformation that relates BMN operators in ${\cal
N}\!=4$ gauge theory to string states in the dual string field
theory at finite $g_2\! =\! J^2/N$. In this basis, the supercharge
truncates at linear order in $g_2$, and the mixing amplitude
between 1 and 2-string states precisely matches with the
(corrected) answer of hep-th/0206073 for the 3-string amplitude in
light-cone string field theory. Supersymmetry then predicts the
order $g_2^2$ contact term in the string bit Hamiltonian. The
resulting leading order mass renormalization of string states
agrees with the recently computed shift in conformal
dimension of BMN operators in the gauge theory.

\end{titlepage}

\newpage

\newcommand{\gtt}{{{}^{{}_ {>}}}}
\newcommand{\stt}{{{}^{{}_{<}}}}
\newcommand{\gst}{{{}^{{}_{\geq}}}}
\newcommand{\sgt}{{{}^{{}_{\leq}}}}
\newcommand{\gt}{{{}^{ >}}}
\newcommand{\st}{{{}^{<}}}

\newsubsection{Introduction and Philosophy}

The BMN correspondence \cite{bmn} equates type IIB string theory
on a plane wave background with a certain limit of ${\cal{N}}=4$
gauge theory at large R-charge $J$, where $N$ is taken to infinity
while the quantities \be \lambda' = { g_{\rm YM}^2 N \over J^2 },
\qquad g_2 = {J^2 \over N} \ee are held fixed. The proposal is
based on a natural identification between the basis of string
theory states and the basis of gauge theory operators, and between
the light-cone string Hamiltonian $P^-$ and the generator $\Delta$
of conformal transformation in the gauge theory via\footnote{The
parameter $\mu$ can be introduced by performing a boost and serves
merely as a bookkeeping device.} \be \label{ops} {2 \over \mu} P^-
= \Delta - J\, . \ee BMN argued, and it was subsequently confirmed
to all orders in $\lambda'$ \cite{grossone,sz}, that this
identification holds at the level of free string theory
($g_2\! =\! 0$).

This beautiful proposal equates two operators which act on
completely different spaces: the light-cone Hamiltonian $P^-$ acts
on the Hilbert space of string field theory, and allows for the
splitting and joining of strings, while $H \equiv \Delta - J$ acts
on the operators of the field theory, and in general mixes
single-trace operators with double- and higher-trace operators.
Light-cone string field theory in the plane wave background has been
constructed in \cite{sv,svtwo}.  On the field theory side,
a number of impressive papers
\cite{kpss,bn,boston,gursoy,bkpss,gmr,boston2,Eynard:2002df} have
pushed the calculations to higher order in $g_2$ with
the aim of showing that (\ref{ops}) continues to hold, thereby
providing an equality between a perturbative, interacting string
theory and perturbative ${\cal N}\!= 4$ gauge theory. It is clear,
however, that at finite $g_2$ the natural identification between
single string states and single trace operators breaks down. For
example, 1-string states are orthogonal to 2-string states for all
$g_2$, but single-trace operators and double-trace operators are
not. This raises the question how to formulate the BMN
correspondence in the interacting string theory.

In order to prove that two operators in (\ref{ops})
are equal, it is sufficient
to prove that they have the same eigenvalues.  If they do, then
there is guaranteed to exist a unitary transformation between
the spaces on which the two operators act. A basis independent
formulation of the BMN correspondence, therefore, is that the
interacting string field theory Hamiltonian ${2 \over \mu} P^-$
and the gauge theory operator $H$ must have the same eigenvalues.

%By comparing the corresponding eigenvectors, one can then
%uniquely\footnote{If the spectrum is nondegenerate; otherwise $U$
%will be ambiguous, up to an arbitrary unitary transformation
%within each degenerate subspace.} determine the transformation $U$
%between the string field theory basis and the gauge theory basis
%as a function of $g_2$.

While this is the minimum that we are allowed to expect from the
BMN correspondence, we can hope to do better. Light-cone string
field theory, as formulated in \cite{gsb,sv,svtwo}, comes with a
natural choice of basis: this string basis (of single and multiple
strings) is neither the BMN basis (of single and multiple traces)
nor the basis of eigenstates of the light-cone Hamiltonian. But
how do we identify the string basis in the gauge theory?

One guess for the string basis was made in
\cite{boston,boston2}, where it was argued that matrix
elements of $P^-$ between 1- and 2-string states should be equated
with the coefficient of the three-point function of the
corresponding BMN operators, multiplied by the difference in
conformal dimension between the incoming and outgoing states. This
proposal appeared to be supported by the subsequent string field
theory calculation done in \cite{svtwo} (see also
\cite{Chu:2002pd,Huang:2002wf,Huang:2002yt,Kiem:2002xn,Lee:2002rm,Lee:2002vz,Janik:2002bd,deMelloKoch:2002nq}).
It turns out, however,
that the final step of the calculation in \cite{svtwo} suffered
from a minus sign error (which we will correct below), which
renders the alleged confirmation of this proposal invalid.

In this paper we propose a new, specific form for the
transformation between the BMN basis and the string basis,
valid to all orders in $g_2$. This basis transformation is trivial
to write down, and has the pleasing feature that it does not
depend on the conformal dimensions of the operators.
%\footnote{In our view, the basis
%transformation should come {\it before} one introduces a
%Hamiltonian.  If the transformation depends on the Hamiltonian, it
%is a sign that one actually has already begun diagonalizing the
%Hamiltonian.}
In fact, our choice of transformation was already identified as a
natural choice in \cite{vv}, where it was shown that all computed
amplitudes in gauge theory are reproduced via a relatively simple
string bit formalism \cite{thornbits,bits,zhou}. 
While most calculations in \cite{vv} were
done in the BMN basis, it was pointed out that there exists a
basis choice with the properties that (i) the inner product is
diagonal, and (ii) the matrix elements of the supersymmetry
generators $Q$ are at most linear in $g_2$ (i.e. $Q$ leads to only
a single string splitting or joining). Here we will show that,
when evaluated in this new basis, the matrix elements of the
string bit Hamiltonian, which via the results of \cite{vv} may be
identified with the gauge theory operator on the right-hand side
of (\ref{ops}), agree precisely with the corrected answer of
\cite{svtwo} for the matrix elements of the continuum string field
theory Hamiltonian $P^-$ appearing on the left-hand side!

The precise match between the three point functions means
that, by combining the two formalisms, we can start filling in
some important questions left open in \cite{sv} and \cite{vv}. A major
technical obstacle in continuum light-cone string field theory is
that higher order contact terms are needed for closure of the
supersymmetry algebra, and that their value (at order $g_2^2$)
affects the leading order shift in the eigenvalues of $P^-$.
However, these contact terms are difficult to compute \cite{progress}.
The supersymmetry algebra of the bit string theory, on the other hand,
is known to all orders in $g_2$ but only to linear order in the
fermions. It appears to be a fruitful strategy, therefore, to make
use of the discretized theory to fix the order $g_2^2$ contact
terms of the continuum theory, while the known non-linear
fermionic form of the continuum interaction vertex may be of
direct help in deriving the complete supersymmetry generators in
the string bit formalism.

\newcommand{\nnn}{{{}_n}}
\newcommand{\mmm}{{{}_m}}

\newsubsection{Identification of the String Basis in Gauge Theory}

${\cal N}\! =  4$ gauge theory in the BMN limit comes with a
natural choice of basis, which coincides with the natural string
basis when $g_2\! = \! 0$: an $n$ string state corresponds to a
product of $n$ single trace BMN operators. We call this basis the
BMN basis, denoted by $|\psi_\nnn\rangle$. At non-zero $g_2$, the
inner product (defined as the overlap as computed in the free
gauge theory) becomes non-diagonal in this basis. The explicit
form of the inner product is conveniently expressed in terms of
the string bit language of \cite{bits,vv} as \be \label{ip}
\langle\psi_\mmm|\psi_\nnn\rangle{}_{g_2}=\, 
\Bigl(e^{g_2 \Sigma}\Bigr)_{nm}  \, , \qquad \quad \Sigma
 = {1\over J^2} \sum_{i<j}\Sigma_{ij}\ee
where $\Sigma_{ij}$ is the operator which interchanges the string
bits via the simple permutation $(ij)$. As explained in
\cite{bits,vv}, when acting on a BMN state $|\psi\rangle$ with $n$
strings, $\Sigma$ effectuates a single string splitting or
joining.

This meaning of $\Sigma$ in the gauge theory language can be made
concrete as follows. Consider a long BMN string in its ground
state. We can write the corresponding operator as \be {\cal
O}_J(\gamma) = {\rm Tr}( Z^{J})= \sum_{{i_1 \ldots i_J} \atop
{\overline{i}_1 \ldots \overline{i}_j}} Z_{i_1 \overline{i}_1}\,
Z_{i_2 \overline{i}_2} \ldots Z_{i_J\overline{i}_J} \;
\delta^{\overline{i}_1 i_{\gamma(1)}}\, \delta^{\overline{i}_2
i_{\gamma(2)}} \ldots \delta^{\overline{i}_J i_{\gamma(J)}} \ee
with $\gamma = (1 2 \ldots J)$ the cyclic permutation of $J$
elements. The action of $\Sigma_{J_1J}$, which implements the
simple permutation $(J_1J)$, is now defined as \be \Sigma_{J_1J}\,
{\cal O}_J(\gamma) = {\cal O}_J(\gamma \circ (J_1J)) \ee Since
$\gamma \circ (J_1J) = (1 \ldots J_1\!-\!1\, J)(J_1\ldots J-1)$ we
have that \be \Sigma_{J_1 J}\, {\cal O}_J(\gamma) = {\rm Tr}
(Z^{J_1} ) {\rm Tr}(Z^{J-J_1}), \ee showing that the simple
permutation $\Sigma_{J_1J}$ indeed induces a single splitting of a
single trace into a double trace operator. It is easy to
generalize this result to other operators, to show that $\Sigma$
can either split a string or join two strings.

The identification of (\ref{ip}) with the inner product of the
free gauge theory was motivated in \cite{vv} and explicitly
verified for string ground states to all order in $g_2$ and for
two-impurity states to order $g_2^2$.

States with different number of strings are therefore no longer
orthogonal relative to (\ref{ip}). In the string field theory
basis $|\tilde{\psi}_\nnn\rangle$, on the other hand, the inner
product should be diagonal for all $g_2$. The simplest basis
transformation that achieves this goal is \be \label{redef}
|\widetilde{\psi}_\nnn\rangle = (e^{-g_2 \Sigma/2})_{\nnn
\mmm}|\psi_\mmm\rangle \, . \ee This is not the most general
diagonalization, however, since we still have the freedom to
redefine the new basis $|\widetilde\psi_\mmm\rangle$ via an
arbitrary unitary transformation \cite{gmr}. The above
redefinition (\ref{redef}), however, has the attractive feature
that it is purely combinatoric and does not depend on the dynamics
of the gauge theory. Furthermore, as we will see shortly, it has
the desirable property that the (linearized) supersymmetry
generators and light-cone Hamiltonian acquire a simple form in the
new basis. We emphasize that the only way to check the proposal
(\ref{redef}) for identifying the string field theory basis in the
gauge theory is by comparing matrix elements of $H$ calculated in
the $|\widetilde{\psi}_n\rangle$ basis to those of ${2 \over \mu}
P^-$ in light-cone string field theory. We show below that the
proposal (\ref{redef}) passes this test.

In the following, we will study the consequences of this basis
transformation for the specific class of two-impurity BMN states
investigated in \cite{kpss,bkpss,boston,boston2,gmr}. We will
denote by $|1,p\rangle$ the normalized state corresponding to the
single trace operator $\sum_l e^{2\pi ipl/J} {\rm Tr}(\phi Z^l
\psi Z^{J-l})$, while $|2,k,y\rangle$ and $|2,y\rangle$ will
denote the normalized states corresponding to the double trace
operators $\sum_l e^{2\pi ikl/{J_1}} {\rm Tr}(\phi Z^l\psi
Z^{J_1-l}){\rm Tr}(Z^{J-J_1})$ and ${\rm Tr}(\phi Z^{J_1}){\rm
Tr}(\psi Z^{J-J_1})$ respectively, where $y = J_1/J$. The action
of $\Sigma$ on $|1,p\rangle$ reads
 \be \label{ope} \Sigma
\, | 1, p \rangle = \sum_{k,y} C_{pky} \, |2, k,y\rangle\;  + \;
\sum_y C_{py} |2,y\rangle \, , \ee with \be \label{cs} C_{pky} =
\sqrt{1 - y \over J y} {\sin^2(\pi p y) \over \pi^2 (p-k/y)^2},
\qquad C_{py} = - {\sin^2(\pi p y) \over \sqrt{J} \pi^2 p^2}. \ee
Via (\ref{redef}) we now introduce the corresponding two-impurity
states in the string basis, which we will denote by $|\wilde{1},
p, y\rangle$, $|\wilde{2},k,y\rangle$ and $|\wilde{2}, y\rangle$,
respectively. By construction, these form an orthonormal basis at
finite $g_2$.

\newsubsection{Interactions in the String Basis}

In this section we obtain the matrix elements of the right-hand
side of (\ref{ops}) in the string basis proposed in the previous
section. For this we will employ the string bit model of
\cite{vv}, but by virtue of the established correspondence with
the gauge theory amplitudes of \cite{bkpss,boston2}, the following
calculation can also be viewed as a direct calculation within the
gauge theory.

It was shown in \cite{vv} that the linearized (in the fermions)
interacting supercharges in the string bit model can be written in
the string basis as \be \label{nnnsusy} Q = Q_\nul + {g_\two\over
2} [ \widehat{Q}_\nul,\Sigma]\, , \qquad \qquad \widehat{Q}_\nul =
Q^\stt_\nul\!\!-\!Q_\nul^\gtt \, , \ee
%\be H = H_0 + H_1 + H_2 \ee
where $Q_\nul = Q^\stt_\nul + Q^\gtt_\nul$ is the free supercharge
of the bit string theory and the superscripts indicate the
projection onto the term with fermionic creation ($<$) or
annihilation ($>$) operators only. These charges generate the
interacting superalgebra of string theory in the plane wave
background, modulo higher order terms in the fermions. Our
interest is to compute the matrix elements of $H$ between the
bosonic two-impurity states in the string basis.

\figuur{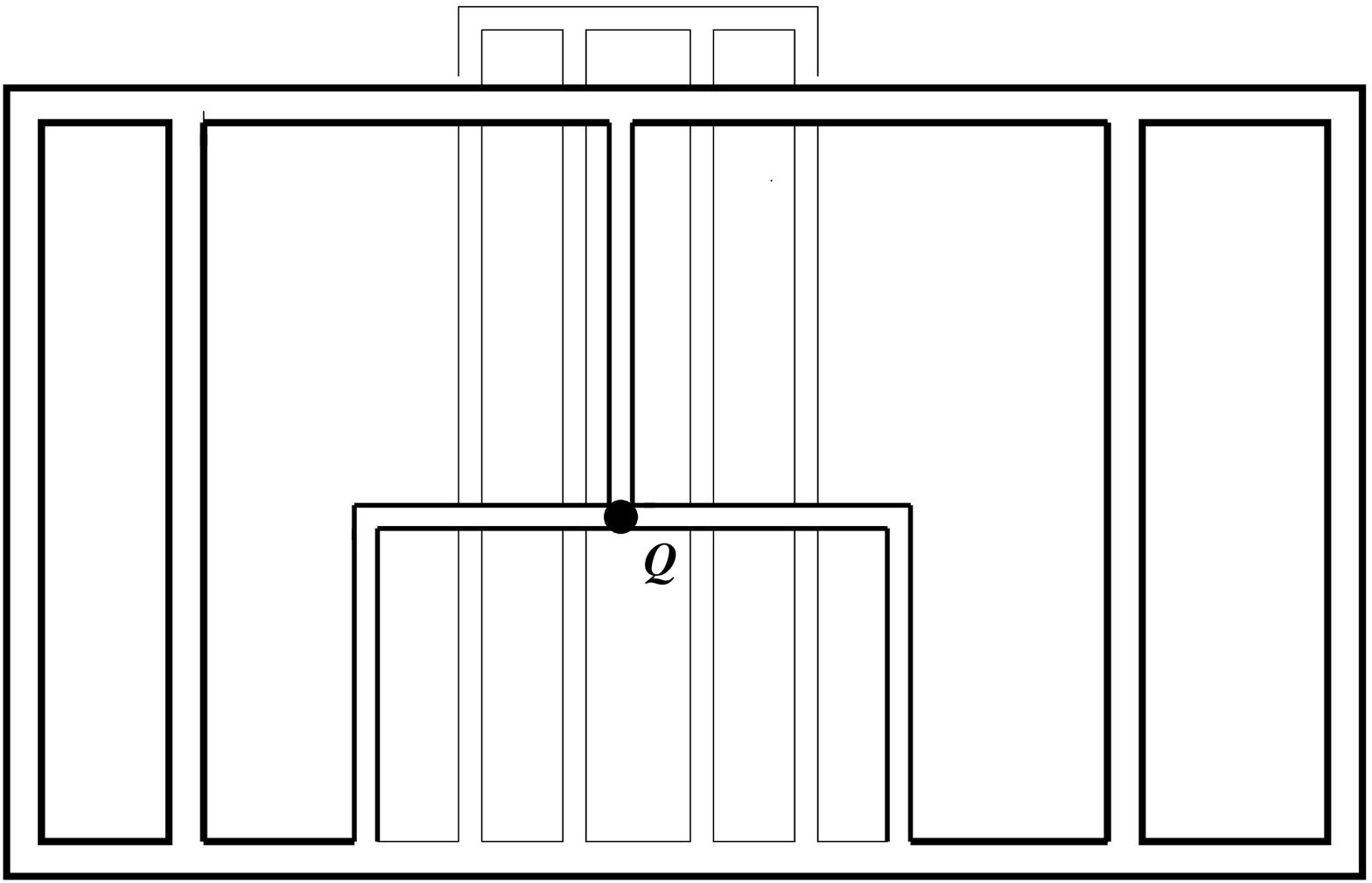}{8cm}{Fig 1. An insertion of the supercharge in the
gauge theory will lead to a single splitting of the BMN string
with which it has a double contraction.}

The supercharge (\ref{nnnsusy}) truncates at linear order in
$g_2$. As indicated in fig 1, this truncation is expected from the
BMN correspondence: matrix elements of the supercharges $Q=\Tr
\theta [Z,\phi]$ in the gauge theory can lead (for connected
diagrams, where $Q$ has at least one contraction with either the
``in'' or ``out'' BMN state) to at most one single string
splitting or joining \cite{vv}. For now, however, one may view
(\ref{nnnsusy}) as a new starting point of the string bit model;
in the remainder we will re-establish its equivalence with the
perturbative gauge theory, by showing that, at least for the
special class of two-impurity states, it leads to the same order
$g_2^2$ mass renormalization (shift in conformal dimension) as
computed in \cite{bkpss,boston2}.

{}From (\ref{nnnsusy}) we thus deduce that the interacting
Hamiltonian truncates at order $g_2^2$: \be \vspace{-3mm} H = H_0
+ g_\two H_1 + g_\two^2 H_2 \ee with\footnote{The notation in
these equations is somewhat symbolic: the left-hand side in each
equation is equal to the projection onto the $\delta^{\da\db}$
component of the anti-commutator on the right-hand side.
Furthermore, as stated above, this formula for the Hamiltonian is
valid only for computing matrix elements between bosonic states;
for fermionic states, the non-linear fermionic corrections to
(\ref{nnnsusy}) will become relevant.}
%\vspace{-3mm}
\be H_0 = \{ Q_\nul, Q_\nul \} \, , \ \ \qquad H_1 = \{ Q_\nul ,
[\widehat{Q}_\nul ,\Sigma]\} \, , \ \qquad  H_2 = {1\over 4} \{
[\widehat{Q}_\nul ,\Sigma], [\widehat{Q}_\nul ,\Sigma]\}\, . \ee
{}From these expressions, it is straightforward to compute the
matrix elements between the various two-impurity states. Using
that $Q^\gtt_\nul$ annihilates bosonic states, we find that \be
\langle \widetilde{\psi}_2 | H_1 | \widetilde{\psi}_1 \rangle = {1
\over 2} \langle \widetilde{\psi}_2 | (H_\nul \Sigma + \Sigma
H_\nul) | \widetilde{\psi}_1 \rangle - 2 \langle \widetilde{\psi}_2|
{Q}^\gtt_\nul \, \Sigma\, Q_\nul^\stt| \widetilde{\psi}_1\rangle.
\ee Both matrix elements on the right-hand side have been computed
in \cite{vv}, with the result \ba \label{triv} \langle \wilde{2},
k, y | ( H_\nul \Sigma + \Sigma H_\nul ) |\wilde{1},p \rangle \!
\is\! \lambda' (p^2 + k^2 /y^2) \,
C_{p ky}\, ,%\langle 2\, k, y|\Sigma| 1, p\rangle\, .
\qquad \\[2.5mm]
\label{announce} \langle \wilde{2}, k, y |\, {Q}^\gtt_\nul \!
\Sigma\, Q_\nul^\stt | \wilde{1}, p\rangle \is  {\lambda'\over 2} \, (p k/y)
\, C_{p ky}\, ,
%\langle 2\, k, y|\Sigma| 1, p\rangle,
\ea where $C_{pky}$ are the three point functions defined in eqn
(\ref{cs}). Inserting the explicit expressions, we find \be
\label{ftresults} \langle \wilde{2},k,y| H_1 | \wilde{1},p\rangle
= {\lambda' \over 2} \sqrt{1 - y \over J y} {\sin^2(\pi p y) \over
\pi^2}, \qquad \langle \wilde{2},y | H_1 | \wilde{1},p\rangle =  -
{\lambda' \over 2} {1 \over \sqrt{J}} {\sin^2(\pi p y) \over
\pi^2}. \ee In a similar way one can obtain the order $g_2^2$
matrix elements between single string states.  We postpone this
discussion to later, and turn now to the continuum string field
theory.

\newsubsection{Light-Cone String Field Theory}

We now investigate the matrix elements of the left-hand side of
(\ref{ops}) in the continuum string theory, in order to compare
with the gauge theory results of the previous section. In
light-cone string field theory the cubic interaction is
conveniently represented as a state in the three-string Hilbert
space.  If we restrict our attention to string states which have
no fermionic excitations, this state can be expressed as \be
\label{pminus} {2 \over \mu} |P^-_1\rangle = - {y(1-y) \over 2}
{\cal{P}} |V\rangle. \ee Here $|V\rangle$ is a squeezed state in
the 3-string Hilbert space, \be \label{vertex} |V\rangle = \exp
\left[ {1 \over 2} \sum_{r,s=1}^3 \sum_{m,n=-\infty}^\infty
a_{m(r)}^{I \dagger} \overline{N}^{(rs)}_{mn} a_{n(s)}^{I \dagger}
\delta_{IJ} \right] |0\rangle, \ee and the prefactor ${\cal{P}}$
is given by
%\be \label{prefactor}
%{\cal{P}} = a_k
%\sum_{r,s=1}^3 \sum_{m,n=-\infty}^\infty
%(U_{m(r)} U_{n(s)} - 1) F_{m(r)} F_{n(s)} a_{m(r)}^{I \dagger}
%a_{n(s)}^{J \dagger}
%v_{IJ},
%\ee
\be \label{prefactor} {\cal{P}} =  \sum_{r=1}^3
\sum_{m=-\infty}^\infty {\omega_{m(r)} \over \mu \alpha_{(r)}}
a_{m(r)}^{I \dagger} a^J_{-m(r)} v_{IJ}, \ee where $v_{IJ} = {\rm
diag}(1_4,-1_4)$, $\omega_{m(r)} = \sqrt{m^2 + \mu^2
\alpha_{(r)}^2}$, and $\alpha_{(r)} = \alpha' p^+_{(r)}$, with the
convention that $\alpha_{(r)}$ is negative for incoming strings 1
and 2 and positive for the outgoing string 3. In order to write
${\cal{P}}$ in this form, we have employed a very useful
factorization identity derived in \cite{schwarz,ari}.  The sign
error in the original version of \cite{svtwo} amounts, after
tracing through some changes of basis, to replacing the $a_{-m}$
by $a_{m}$ in (\ref{prefactor}), making manifest the incorrect
claim that the prefactor gives the difference of energy between
incoming and outgoing states.\footnote{The error can also be 
understood as a missing factor of $i$ in eqn (3.15) of \cite{sv},
as pointed out in \cite{ari}.}

Implicit formulas for the matrix elements of $\overline{N}$,
valid for all $\lambda'$, were presented in \cite{sv}.
While explicit formulas for the leading terms in an expansion
around $\lambda' = 0$ are known,
it is very difficult to extract exact formulas for higher-order
terms \cite{ksv}.
At $\lambda' = 0$ the only nonzero Neumann matrices are
\ba \label{nlargemu}
\overline{N}^{(13)}_{kp} \is {(-1)^{k+p+1} \over \sqrt{y}}
{\sin (\pi p y) \over \pi (p - k/y)},\\[3.5mm]
\overline{N}^{(23)}_{kp} \is -{1 \over \sqrt{1-y}}
{\sin(\pi p(1- y)) \over \pi (p - k/(1-y))}.
\eea

Using the above formulas, it is straightforward to derive
the leading ${\cal{O}}(\lambda')$ contribution to the matrix elements
\ba \label{sftresults}
{2 \over \mu} \langle \tilde{2},k,y | P^-_1 | \tilde{1},p\rangle \is
{\lambda' \over 2}
(1-y) {\sin^2(\pi p y) \over \pi^2},\\[3.5mm]
{2 \over \mu} \langle \tilde{2},y | P^-_1 |\tilde{1},p \rangle \is
-{\lambda' \over 2} \sqrt{y(1-y)}
{\sin^2(\pi p y) \over \pi^2}. \nonumber
\eea
This result, which corrects the one originally reported in \cite{svtwo},
is in precise agreement with (\ref{ftresults}) after taking into
account the factor $\sqrt{J y(1-y)}$ which arises because (\ref{ftresults})
is written in terms of unit normalized states while
(\ref{sftresults}) is expressed in terms of continuum
states satisfying $\langle i|j \rangle = p_i^+ \delta(p_i^+ - p_j^+)$.

\newsubsection{Contact Terms and Mass Renormalization}

Having established that the matrix elements of $H$ evaluated in
the $|\widetilde{\psi}_\nnn\rangle$ basis agree at order $g_2$
with those of ${2 \over \mu} P^-$ in the natural string field theory basis,
let us now revisit the issues of contact terms and the one-loop
mass renormalization of the single-string state
$|\wilde{1},p\rangle$.

A significant advantage of the $|\widetilde{\psi}\rangle$ basis
is that the matrix elements of the supercharge (\ref{nnnsusy}) terminate
at order $g_2$.  Therefore the Hamiltonian terminates at order $g_2^2$,
with the term $H_2$ which comes from squaring the order $g_2$ term in the
supercharge;
there is no need for all of the higher-order contact terms which seem to plague
continuum light-cone string field theory.  Turning this observation around
gives a definite prediction for string field theory: that the only contact
term surviving in the large $\mu$ limit is the one which comes from squaring
the cubic vertex in the dynamical supercharge.

The order $g_2^2$ matrix element between single string states has
been computed in \cite{vv} (see equation (71)), with the result \ba
\label{contact} g_2^2 \langle \wilde{1},q | H_2 | \wilde{1},p
\rangle \is {g_2^2 \over 4} \langle \wilde{1}, q  | \, [\,
{Q}_\nul^\gtt\! , \Sigma ] [\,
\Sigma , Q_\nul^\stt]\, | \wilde{1}, p \, \rangle \nonumber \\[3mm]
\is
{g_2^2 \lambda'\over 4} \sum_i (k^2/y^2-{1\over 2}(p^2+q^2))C_{p\, i}C_{qi}\
\equiv
{1 \over 4}
 {g_2^2 \lambda'\over 4 \pi^2} \, B_{pq}.
\ea Here the sum runs over intermediate 2-string states $i$ of
both types: $|\wilde{2}, k,y\rangle$ and $|\wilde{2}, y\rangle$
(and includes an integral over $y$).
The explicit form of $B_{pq}$ is as given in
\cite{bkpss,boston2}.

To obtain the order $g_2^2$ mass renormalization of the state
$|\wilde{1},p\rangle$, we should add the matrix element
(\ref{contact}) to the iterated $H_1$ interaction:\footnote{Here
in the second step we use that $C_{-pky} = {(p-k/y)^2\over
(p+k/y)^2}\, C_{pky}$.} \ba \label{itone} \sum_i { |\langle \,
\wilde{1},p\, |\, H_1 | \, i\, \rangle|^2 \over E_p - E_i} \is {1
\over 4} g_2^2 \lambda' \sum_{i} {(p-k/y)^4 C_{pi}^2 \over
p^2 - k^2/y^2}\\[3.5mm]
\is {1 \over 4} g_2^2 \lambda'  \sum_i (p^2 - k^2/y^2) C_{pi}
C_{-pi}\\
\is - {1 \over 4}  {g_2^2 \lambda' \over  4 \pi^2} B_{p,-p}. \eea
An additional subtlety in the calculation is that the states
$|\wilde{1},p\rangle$ and $|\wilde{1},-p\rangle$ are degenerate at
lowest order; we should check therefore whether we need to use
degenerate perturbation theory.  It is easy to see that \be \sum_i
{ \langle \, \wilde{1},p\, |\, H_1 |\, i\, \rangle\langle\, i\,
|\, H_1 |\, \wilde{1},-p\rangle \over E_p - E_i} = - {1 \over 4}
{g_2^2 \lambda' \over 4 \pi^2} B_{p,-p} \ee gives the same result
as (\ref{itone}). The sum of the contact term $H_2$ and the
iterated $H_1$ interaction is diagonal in the
$\{|\wilde{1},p\rangle, |\wilde{1},-p\rangle \}$ basis, signalling
that the degeneracy remains unbroken to this order.\footnote{In
fact, supersymmetry requires these two states to be exactly
degenerate \cite{boston2}.  The fact that we find degeneracy
at this order is consistent with our observation that no
additional contact terms
are required for closure of the supersymmetry algebra.}

Putting everything together, we find that the order $g_2^2$ contribution
to the eigenvalue is \be {1 \over 4} {g_2^2 \lambda' \over 4 \pi^2}
(B_{pp} - B_{p,-p}) = {g_2^2 \lambda' \over 4 \pi^2} \left( {1
\over 12} + {35 \over 32 \pi^2 p^2}\right). \ee This agrees precisely
with the shift in anomalous dimension of the conformal
eigen-operators, as reported in
\cite{bkpss,boston2}.

\newsubsection{Conclusion}

We have proposed an explicit form (\ref{redef})
of the basis transformation that
relates single and multi-trace BMN operators in the gauge theory
to single and multi-string states of the dual string field theory in the
plane wave background to all orders
in $g_2$. This basis transformation is natural from
the point of view of the bit string theory of \cite{vv}: besides
the fact that it diagonalizes the inner product, it has the
property that the supersymmetry generators in the new basis
truncate at linear order in $g_2$.

Our most encouraging result, however, is that the 3-point function
in this basis precisely matches with the 3-string amplitude of the
continuum string field theory \cite{svtwo}. In itself this match
does not yet prove anything, because one can always find two bases
that would lead to the same 3-point function. One also needs
control over the order $g_2^2$ contact interactions before one can
honestly compare the shift in the conformal dimensions in the
gauge theory with the mass renormalization in the string theory
\cite{progress}. However, we have more information than just the
3-point function: because the supersymmetry charge of the bit
string model is linear in $g_2$, via the closure of the
supersymmetry algebra we have a principle that uniquely determines
the order $g_2^2$ contact interaction. It also tells us that any
higher order contact terms are absent.

It would clearly be of interest to give a precise construction of
the continuum limit of the bit string theory by taking the large
$J$ limit while keeping $\lambda'$ fixed. The correspondence found
in this paper is an encouraging indication that this continuum
limit will coincide with the continuum light-cone string theory.

\bigskip

\newsubsection{Acknowledgements}

It is a pleasure to thank D. Freedman, D. Gross, M. Headrick,
S. Minwalla,
L. Motl, R. Roiban, J. Schwarz and G. Semenoff
for useful discussions.
This research was supported in part by NSF grants PHY98-02484 (JP, DV, HV)
and PHY99-07949 (AV),
DOE grant
DE-FG02-91ER40671 (MS), and an NSF Graduate
Research Fellowship (JP).
MS and AV would also like to thank the Aspen Center for Physics for
hospitality during the early stages of this work.
Any opinions, findings, and conclusions or recommendations expressed in
this material are those of the authors and do not necessarily reflect
the views of the National Science Foundation.

%%%%%%%%%%%%%%%%%%%%%%%%%%%%%%%%%%%%%%%%%%%%%%%%%

%%%%%%%%%%%%%%%%%%%%%%%%%%%%%%%%%%%%%%%%%%%%%


\begin{thebibliography}{30}
\parskip-2pt
%%%%%%%%%%%%%%%%%%%%%%%%%%%%%%%%%%%%%%%%%%%%%%%%%%%%%%%

\bibitem{bmn}
D.~Berenstein, J.~M.~Maldacena and H.~Nastase,
``Strings in flat space and pp waves from N = 4 super Yang Mills,''
JHEP {\bf 0204}, 013 (2002)
[arXiv:hep-th/0202021].
%%CITATION = HEP-TH 0202021;%%

\bibitem{grossone}
D.~J.~Gross, A.~Mikhailov and R.~Roiban,
``Operators with large R charge in N = 4 Yang-Mills theory,''
Annals Phys.\  {\bf 301}, 31 (2002)
[arXiv:hep-th/0205066].
%%CITATION = HEP-TH 0205066;%%

\bibitem{sz}
A.~Santambrogio and D.~Zanon,
``Exact anomalous dimensions of N = 4 Yang-Mills operators with large R
charge,''
arXiv:hep-th/0206079.
%%CITATION = HEP-TH 0206079;%%

\bibitem{sv}
M.~Spradlin and A.~Volovich,
``Superstring interactions in a pp-wave background,''
arXiv:hep-th/0204146.
%%CITATION = HEP-TH 0204146;%%

\bibitem{svtwo}
M.~Spradlin and A.~Volovich,
``Superstring interactions in a pp-wave background. II,''
arXiv:hep-th/0206073.
%%CITATION = HEP-TH 0206073;%%

\bibitem{gsb}
M.~B.~Green, J.~H.~Schwarz and L.~Brink,
``Superfield Theory Of Type II Superstrings,''
Nucl.\ Phys.\ B {\bf 219}, 437 (1983).
%%CITATION = NUPHA,B219,437;%%

\bibitem{kpss}
C.~Kristjansen, J.~Plefka, G.~W.~Semenoff and M.~Staudacher,
``A new double-scaling limit of N = 4 super Yang-Mills theory and PP-wave
strings,''
Nucl.\ Phys.\ B {\bf 643}, 3 (2002)
[arXiv:hep-th/0205033].
%%CITATION = HEP-TH 0205033;%%

\bibitem{bn}
D.~Berenstein and H.~Nastase,
``On lightcone string field theory from super Yang-Mills and holography,''
arXiv:hep-th/0205048.
%%CITATION = HEP-TH 0205048;%%

\bibitem{boston}
N.~R.~Constable, D.~Z.~Freedman, M.~Headrick, S.~Minwalla, L.~Motl,
A.~Postnikov and W.~Skiba,
``PP-wave string interactions from perturbative Yang-Mills theory,''
JHEP {\bf 0207}, 017 (2002)
[arXiv:hep-th/0205089].

\bibitem{gursoy}
U.~Gursoy,
``Vector operators in the BMN correspondence,''
arXiv:hep-th/0208041.
%%CITATION = HEP-TH 0208041;%%

\bibitem{bkpss}
N.~Beisert, C.~Kristjansen, J.~Plefka, G.~W.~Semenoff and M.~Staudacher,
``BMN correlators and operator mixing in N = 4 super Yang-Mills theory,''
arXiv:hep-th/0208178.
%%CITATION = HEP-TH 0208178;%%

\bibitem{gmr}
D.~J.~Gross, A.~Mikhailov and R.~Roiban,
``A calculation of the plane wave string Hamiltonian from N = 4
super-Yang-Mills theory,''
arXiv:hep-th/0208231.
%%CITATION = HEP-TH 0208231;%%

\bibitem{boston2}
N.~R.~Constable, D.~Z.~Freedman, M.~Headrick and S.~Minwalla,
``Operator mixing and the BMN correspondence,''
arXiv:hep-th/0209002.
%%CITATION = HEP-TH 0209002;%%

\bibitem{Eynard:2002df}
B.~Eynard and C.~Kristjansen,
``BMN correlators by loop equations,''
arXiv:hep-th/0209244.
%%CITATION = HEP-TH 0209244;%%

\bibitem{thornbits}
%Giles:1977mp}
{R.~Giles and C.~B.~Thorn,
``A Lattice Approach To String Theory,''
Phys.\ Rev.\ D {\bf 16}, 366 (1977);
%%CITATION = PHRVA,D16,366;%%
C.~B.~Thorn, {
``Supersymmetric quantum mechanics for
string-bits,''} Phys.\ Rev.\ D {\bf 56}, 6619 (1997)
[arXiv:hep-th/9707048];
%%CITATION = HEP-TH 9707048;%%
A related approach to large $N$ gauge theory has been formulated
in:
%\cite{Thorn:1979gu}
C.~B.~Thorn,
``A Fock Space Description Of The 1/N-C Expansion Of Quantum Chromodynamics,''
Phys.\ Rev.\ D {\bf 20}, 1435 (1979).}
%%CITATION = PHRVA,D20,1435;%%

\bibitem{bits}
H.~Verlinde,
``Bits, matrices and 1/N,''
arXiv:hep-th/0206059.
%%CITATION = HEP-TH 0206059;%%
%\cite{Zhou:2002mi}
\bibitem{zhou}
J.~G.~Zhou, ``PP-wave string interactions from string bit model,''
arXiv:hep-th/0208232.
%%CITATION = HEP-TH 0208232;%%

\bibitem{vv}
D.~Vaman and H.~Verlinde,
``Bit strings from N = 4 gauge theory,''
arXiv:hep-th/0209215.
%%CITATION = HEP-TH 0209215;%%

\bibitem{progress}
R.~Roiban, M.~Spradlin and A.~Volovich, in progress.

\bibitem{Kiem:2002xn}
Y.~j.~Kiem, Y.~b.~Kim, S.~m.~Lee and J.~m.~Park,
``pp-wave / Yang-Mills correspondence: An explicit check,''
Nucl.\ Phys.\ B {\bf 642}, 389 (2002)
[arXiv:hep-th/0205279].
%%CITATION = HEP-TH 0205279;%%

\bibitem{Huang:2002wf}
M.~x.~Huang,
``Three point functions of N = 4 super Yang Mills from light cone string
field theory in pp-wave,''
Phys.\ Lett.\ B {\bf 542}, 255 (2002)
[arXiv:hep-th/0205311].
%%CITATION = HEP-TH 0205311;%%

\bibitem{Chu:2002pd}
C.~S.~Chu, V.~V.~Khoze and G.~Travaglini,
``Three-point functions in N = 4 Yang-Mills theory and pp-waves,''
JHEP {\bf 0206}, 011 (2002)
[arXiv:hep-th/0206005].
%%CITATION = HEP-TH 0206005;%%

\bibitem{Lee:2002rm}
P.~Lee, S.~Moriyama and J.~w.~Park,
``Cubic interactions in pp-wave light cone string field theory,''
arXiv:hep-th/0206065.
%%CITATION = HEP-TH 0206065;%%

\bibitem{Huang:2002yt}
M.~x.~Huang,
``String interactions in pp-wave from N = 4 super Yang Mills,''
arXiv:hep-th/0206248.
%%CITATION = HEP-TH 0206248;%%

\bibitem{Lee:2002vz}
P.~Lee, S.~Moriyama and J.~w.~Park,
``A note on cubic interactions in pp-wave light cone string field  theory,''
arXiv:hep-th/0209011.
%%CITATION = HEP-TH 0209011;%%

\bibitem{deMelloKoch:2002nq}
R.~de Mello Koch, A.~Jevicki and J.~P.~Rodrigues,
``Collective string field theory of matrix models in the BMN limit,''
arXiv:hep-th/0209155.
%%CITATION = HEP-TH 0209155;%%

\bibitem{Janik:2002bd}
R.~A.~Janik,
``BMN operators and string field theory,''
arXiv:hep-th/0209263.
%%CITATION = HEP-TH 0209263;%%

\bibitem{schwarz}
J.~H.~Schwarz,
``Comments on superstring interactions in a plane-wave background,''
arXiv:hep-th/0208179.
%%CITATION = HEP-TH 0208179;%%





\bibitem{ari}
A.~Pankiewicz,
``More comments on superstring interactions in the pp-wave background,''
arXiv:hep-th/0208209.
%%CITATION = HEP-TH 0208209;%%

\bibitem{ksv}
I.~R.~Klebanov, M.~Spradlin and A.~Volovich,
``New effects in gauge theory from pp-wave superstrings,''
arXiv:hep-th/0206221.
%%CITATION = HEP-TH 0206221;%%

\bibitem{Parnachev:2002kk}
A.~Parnachev and A.~V.~Ryzhov,
``Strings in the near plane wave background and AdS/CFT,''
arXiv:hep-th/0208010.
%%CITATION = HEP-TH 0208010;%%

\end{thebibliography}
\end{document}